\documentclass[doublecol]{epl2} 

\usepackage{amssymb}
\usepackage{amsmath}
\usepackage{graphicx}
\usepackage{amsbsy}
\usepackage{color}
\usepackage{bm}

\newcommand{\bq}{\begin{eqnarray}}
\newcommand{\eq}{\end{eqnarray}}
\newcommand{\bqn}{\begin{eqnarray*}}
\newcommand{\eqn}{\end{eqnarray*}}

\newcommand{\rr}{\mathbf{r}}

\newcommand{\bwt}{\begin{widetext}}
\newcommand{\ewt}{\end{widetext}}

\title{{{The} restricted primitive model of ionic fluids with nonadditive diameters}}

\author{Riccardo Fantoni\inst{1} \and Giorgio Pastore\inst{2}}
\institute{
\inst{1} Dipartimento di Scienze dei Materiali e Nanosistemi,
  Universit\`a Ca' Foscari Venezia, Calle Larga S. Marta DD2137, I-30123
  Venezia, Italy\\
\inst{2} Dipartimento di Fisica dell' Universit\`a and
  IOM-CNR, Strada Costiera 11, 34151 Trieste, Italy
}

\abstract{
The restricted primitive model with nonadditive hard-sphere diameters
is shown to have interesting and peculiar clustering properties. We
report accurate calculations of the cluster
concentrations. Implementing efficient and ad hoc Monte Carlo
algorithms we determine the effect of nonadditivity on both the
clustering and the gas-liquid binodal. For negative nonadditivity,
tending to the extreme case of completely overlapping unlike ions, the
prevailing clusters are made of an even number of particles having
zero total charge. For positive nonadditivity, the frustrated
tendency to segregation of like particles and the reduced space 
available to the ions favors percolating clusters at high densities.  
}


\begin{document}
\maketitle
\label{sec:introduction}

Ionic soft matter \cite{deGennes1992,Henderson2004} is a class of
conventional condensed soft matter whose interactions are dominated by
electrostatics crucially affecting its physical properties. Among the
most popular representatives of such a class of materials are natural
and synthetic saline environments, like aqueous and non-aqueous
electrolyte solutions and molten salts, including room-temperature
ionic liquids, as well as a variety of polyelectrolytes and colloidal 
suspensions. Equally well known are biological systems of
proteins. 

The simplest fluid modeling an ionic colloidal suspension is the
Restricted Primitive Model (RPM) \cite{Hansen} a binary mixture of
uniformly charged Hard-Spheres (HS) {for which the
  like-unlike collision diameter between a particle of species $1$, of
  diameter $\sigma_{11}=\sigma$, and a particle of species $2$ of
  diameter $\sigma_{22}=\sigma$, is equal to the arithmetic mean
  $\sigma_{12}^\text{add}= (\sigma_{11}+\sigma_{22})/2=\sigma$ {.
  The} two species are of charge $\pm q$ with equal concentrations to
  ensure charge neutrality, and the particles 
move} in a medium of fixed dielectric constant $\epsilon$. The
phase diagram of this model has been widely studied both within
computer experiments 
\cite{Orkoulas1994,Caillol1995,Orkoulas1999,Camp1999,Enrique2000,Luijten2002,Caillol2002}
and through analytical theories \cite{Stell1976,Gillan1983,Given1992,Given1992b,Fisher1993,Zhou1995,Given1997,Jiang2002}.  

From these studies emerged how, in the vapor phase of this fluid, and
thus in the determination of the phase diagram, an important role is
played by association and clustering. In an old paper,
\cite{Pastore1985} one of us studied a {more general} RPM
fluid where it is allowed for size nonadditivity amongst the
particles{: the like-unlike collision diameter differs from
  $\sigma_{12}^\text{add}$ by a quantity
  $\Delta=(\sigma_{12}-\sigma_{12}^\text{add})/\sigma_{12}^\text{add}$
  called the nonadditivity parameter. It was suggested through the use
  of integral equation theories, that such a fluid might have a
  dramatic change of its clustering properties. The nonadditivity of
  the HS diameters does not destroy the simplifying symmetry of the
  model but it introduces modifications of the properties of
the pure RPM model making it a paradigm for the self-assembly of
isotropic particles and a challenge to present day theories of fluids.
There seems to be a lack of literature on this subject
  excepted for Ref. \cite{ZuckermanPhD}.}

In this letter, we reconsider such a model fluid by using more
direct, highly {efficient numerical simulations}. In particular we 
analyze the clustering properties outside of the gas-liquid
coexistence region. As we will see the clustering turns out 
to be greatly affected by the nonadditivity parameter. To the best of
our knowledge this is the first time that such a model fluid is studied
with numerical simulations. The debate on the importance of
clustering in the RPM is rejuvenated by studying this new model
fluid.

\label{sec:model}

The model system here considered may be realized experimentally through a
colloid-star polymer mixture where both species are charged
\cite{Poon2001,Poon2002} {and may be relevant for modeling
  room temperature ionic liquids
\cite{Weingartner1995,Kleemeier1999,Saracsan2006,Schroer2009}}. It is
the restricted primitive model 
(RPM) of nonadditive charged hard-spheres (NACHS). The RPM consists of $N/2$
uniformly charged hard-spheres of diameter $\sigma$ carrying a total
charge $+q$ and $N/2$ uniformly charged hard-spheres of the same
diameter carrying a total charge $-q$. The spheres are moving in a
dielectric continuum of dielectric constant $\epsilon$. The
interaction between ions of apecies $i$ and $j$ a distance $r$ apart
is given by  
\bq
\beta\phi_{ij}(r)=\left\{\begin{array}{ll}
+\infty & r\le \sigma_{ij}\\
\displaystyle
\frac{q_iq_j}{k_BT\epsilon r} & r>\sigma_{ij}
\end{array}\right.~,~~~i,j=1,2~,
\eq
where $\beta=1/k_BT$ with $T$ the absolute temperature and $k_B$ the
Boltzmann's constant, $q_i$ the charge of an ion of species $i$.
The ions form a mixture of NACHS,
i.e. $\sigma_{11}=\sigma_{22}=\sigma$ and $\sigma_{12}=\sigma (1+\Delta)$,
with $\Delta>-1$ the nonadditivity parameter. A thermodynamic state
is completely specified by the reduced density
$\rho^*=\rho\sigma^3=N\sigma^3/V$, where $V$ is the volume containing
the fluid, 
the reduced temperature $T^*=k_BT\epsilon\sigma/q^2$, and the
nonadditivity parameter $\Delta$. 

\label{sec:results}

We used canonical $NVT$ Monte Carlo (MC) simulations to study the
fluid in a cubic simulation box of volume $V=L^3$ with periodic 
boundary
conditions. The long range of the $1/r$ interaction was accounted for
using the  Ewald method \cite{Allen}.

We start from a simple cubic configuration of two crystals one
made of species 1 and one made of species 2 juxtaposed. {The
  maximum particle displacement, the same  along each direction, is
  determined during the first stage of the equilibration run in such
  a way to ensure an average acceptance ratio of $50\%$}. We need around 
$10^5$ MC steps (MCS) in order to equilibrate the samples
and {$10^6$} MCS$/$particle for the statistics.

\label{sec:clusters}

During the simulation we perform a cluster analysis in the vapor
phase. After each 100 
MCS we determine the number $N_n$ of clusters  made of
$n$ particles, so that $\sum_n nN_n=N$. We assume
\cite{Fantoni2011,Fantoni2012} that a group of ions
forms a cluster if the distance $r$, calculated using periodic boundary
conditions, between a particle of species $i$ of the group and
at least one other particle of species $j$ is less then some fixed
value, {\sl i.e.} $r<\sigma_{ij}+\delta\sigma$ where 
$\delta$ is a parameter \footnote{Many different ways of
defining a cluster have been
proposed \cite{Lee1973,Ebeling1980,Gillan1983,Fisher1993,Friedman1979}
since the Bjerrum theory \cite{Bjerrum1926} of ionic associations
first appeared. Our choice corresponds to the geometric one
of Gillan \cite{Gillan1983}.}. {In all our simulations we
  chose $\delta=0.1$ (in Ref. \cite{Caillol1995} a detailed study of
  the sensitivity of the clustering properties on this parameter is
  carried out for the pure RPM fluid).} Then we take the average of
these numbers $\langle N_n\rangle$. 
Here $Q_n=n\langle N_n\rangle/N$ gives the
probability that a particle belongs to a cluster of size $n$. 
{To establish a criterion for percolation, we also find
  clusters without using periodic boundary conditions. One of {\it
    these} clusters percolates if, amongst its particles, there are
  two that do not satisfy the cluster condition {\it as a pair}, but
  do satisfy the condition if periodic boundary conditions are used.}

\begin{figure*}[hbt]
\begin{center}
\includegraphics[width=7cm]{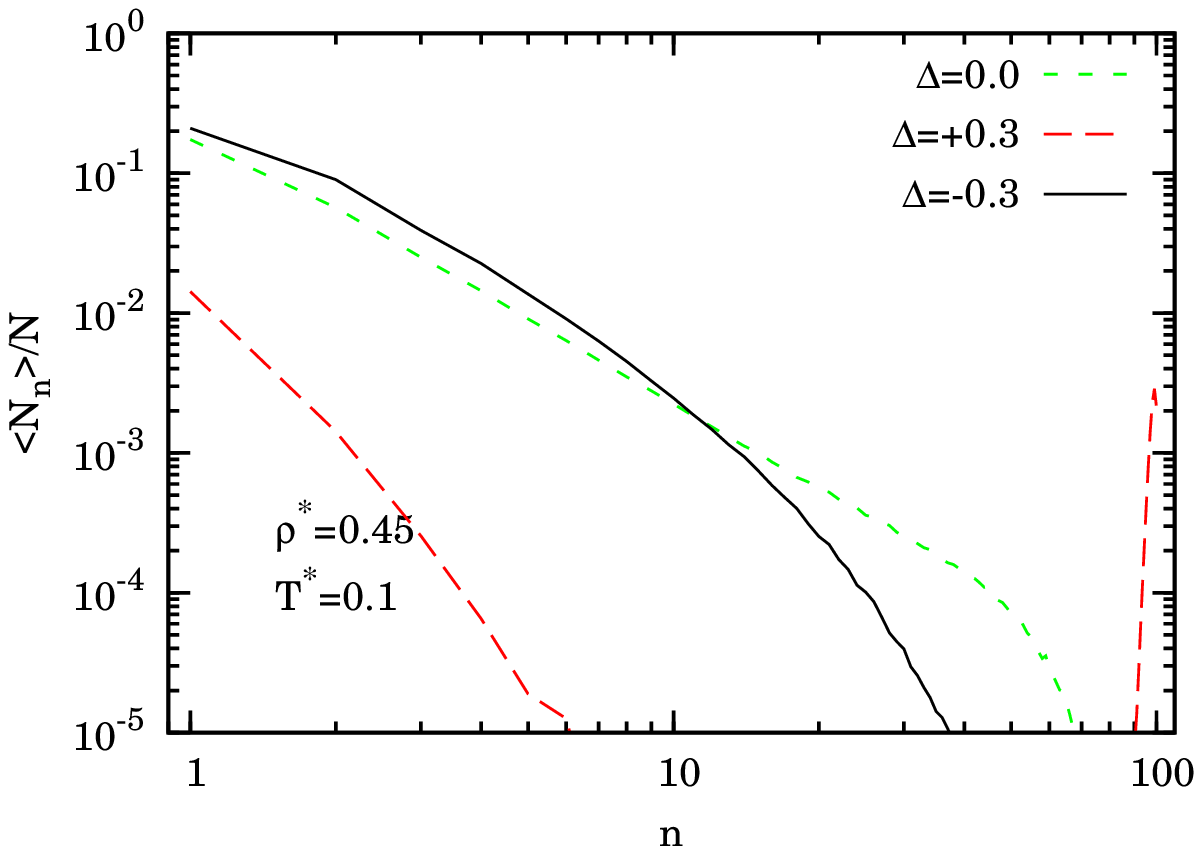}
\includegraphics[width=7cm]{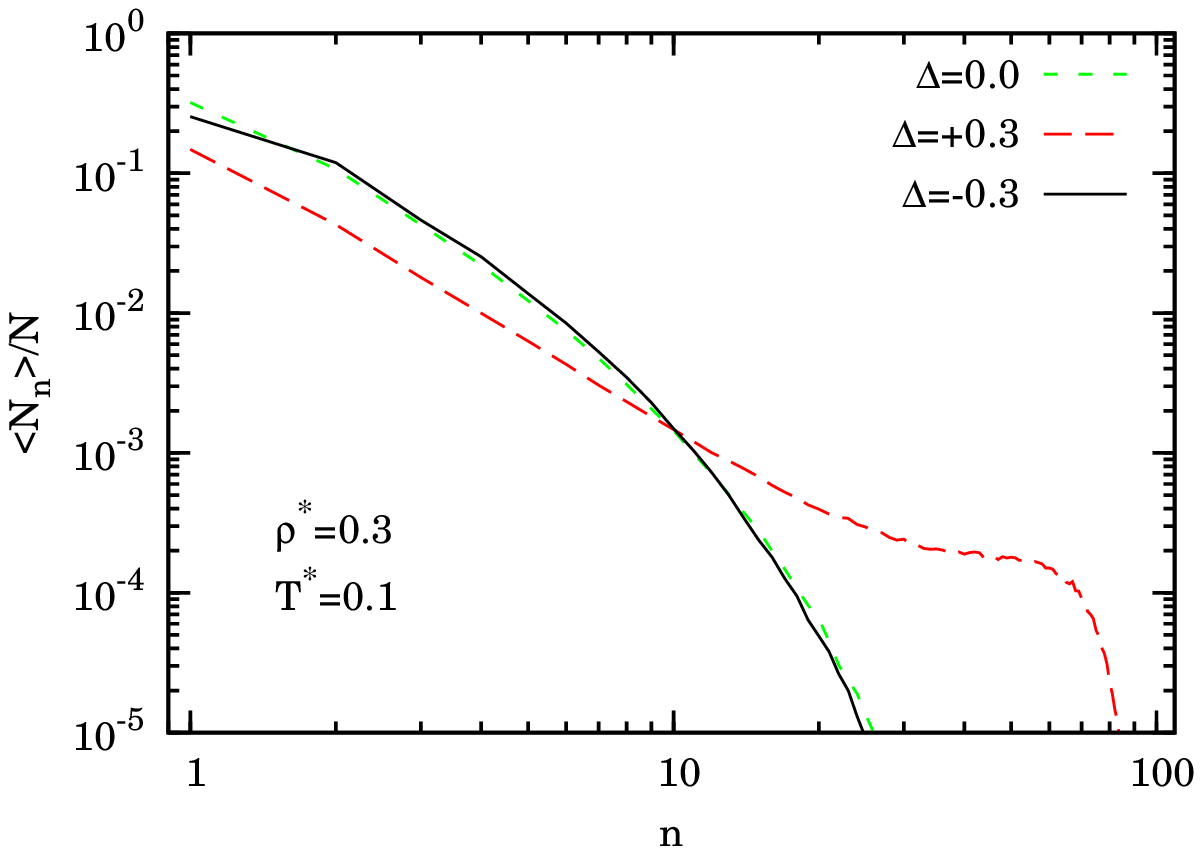}\\
\includegraphics[width=7cm]{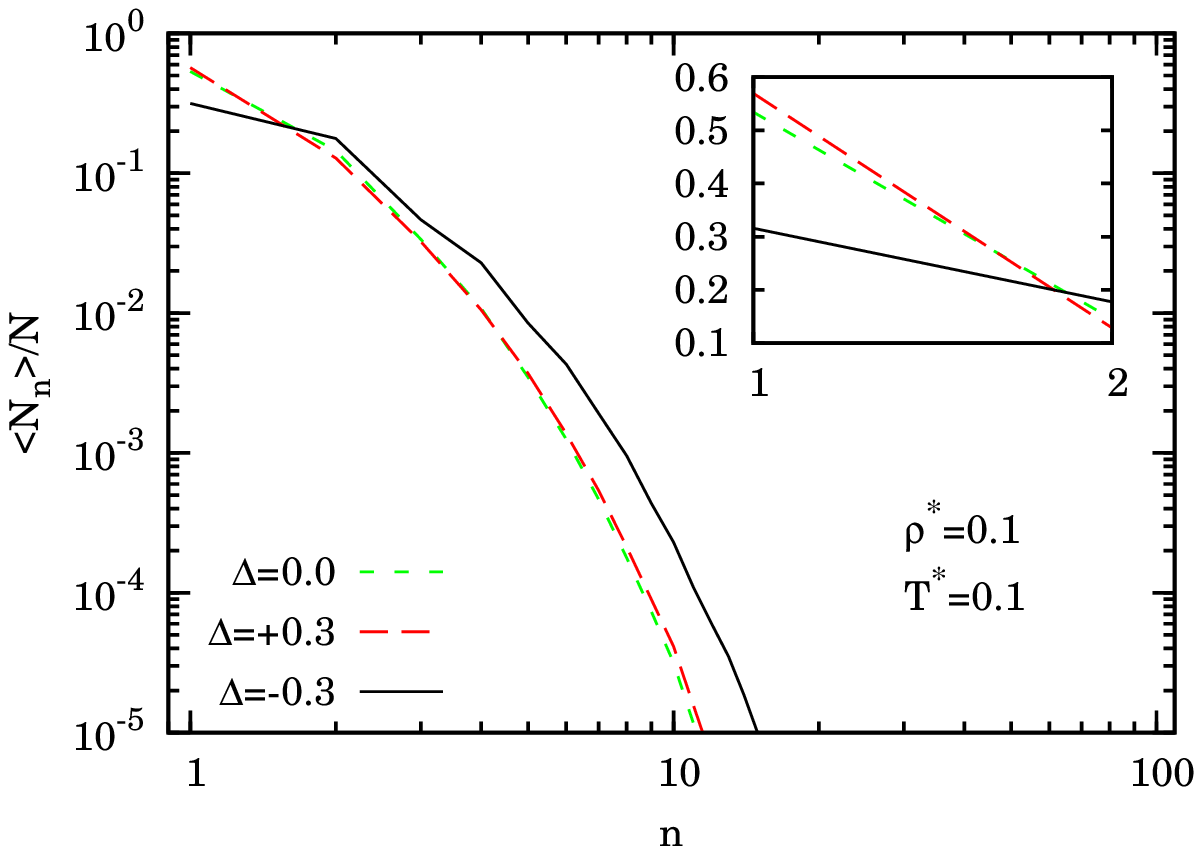}
\includegraphics[width=7cm]{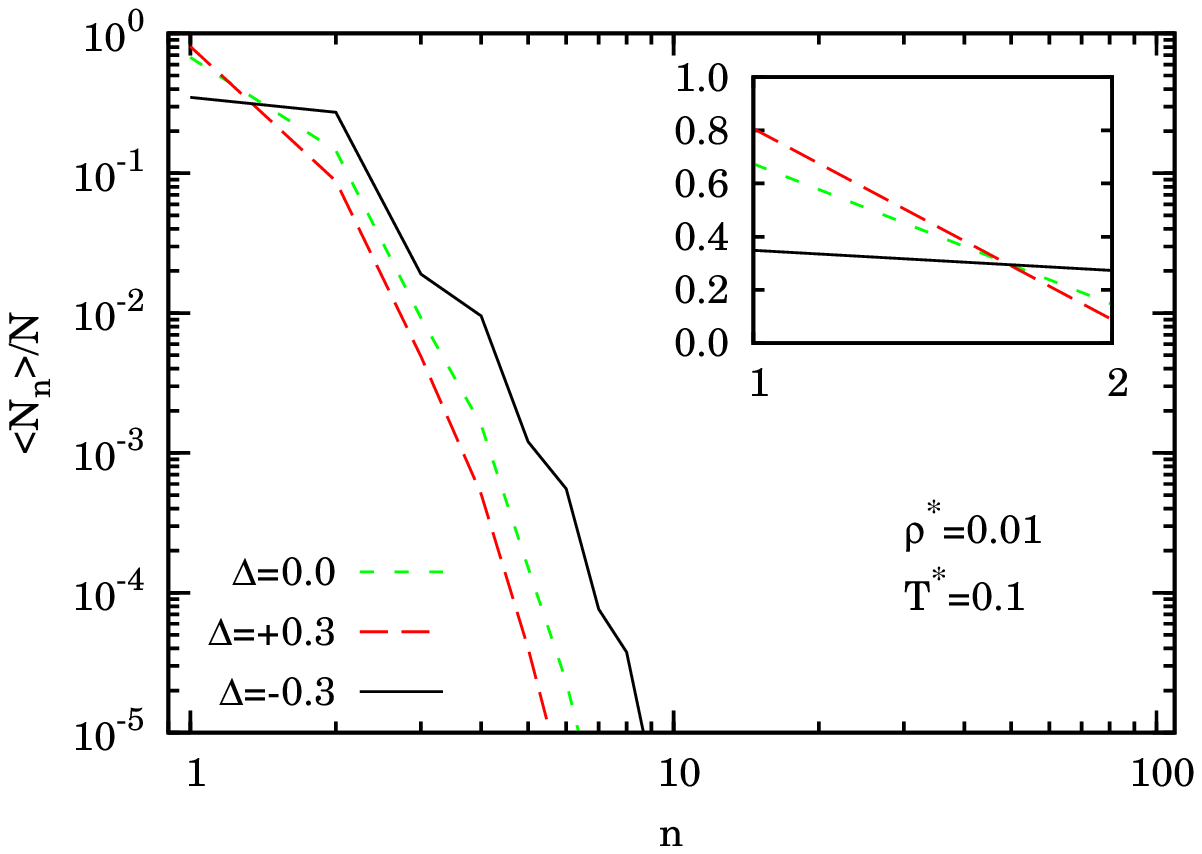}
\end{center}  
\caption{(color online). Clustering properties of the fluid at various
values of nonadditivity and density. $N_n$ are the number
of  clusters made of $n$ particles. We chose
$\delta=0.1$. In the MC simulations we used $N=100$
particles and a number of MCS$=10^7$. {The insets allow 
$\alpha=\langle N_1\rangle/N$, the degree of dissociation, to be
  directly read-off from the graph}.}   
\label{fig:ncl-t0.1}
\end{figure*}

In Fig. \ref{fig:ncl-t0.1} we simulated the fluid at a temperature
$T^*=0.1$ above the critical temperature, $T^*_c\approx 0.05$, of the
pure RPM \cite{Orkoulas1999,Caillol2002,Luijten2002}. 
We see how, at high density, a positive nonadditivity is
responsible for a gain of clustering in the fluid, which  
tends to admit percolating clusters also due to the fact that a
positive nonadditivity 
pushes the fluid at densities closer to the maximum density
attainable. It is well known that in the neutral nonadditive
hard-sphere fluid a positive nonadditivity tends to demix the mixture
at lower densities as $\Delta$ increases 
\cite{Rovere1994,Lomba1996,Jagannathan2003,Gozdz2003,Buhot2005,Santos2010},
so in our fluid we will have a competition between the tendency to
demix in the neutral nonadditive hard-sphere fluid and the tendency to
cluster in the RPM fluid. {At} $\rho^*=0.45$ both 
the pure RPM and the $\Delta=+0.3$ have percolating clusters. Lowering
the density we first reach 
a state, at $\rho^*=0.3$, where the negative nonadditivity gives the
same clustering {as the} RPM and the positive nonadditivity
gives higher 
percolating clustering, then a state, at $\rho^*=0.1$, where
the positive nonadditivity gives the same clustering of RPM and the
negative nonadditivity a higher one, and finally a state, at
$\rho^*=0.01$, at low densities where a negative 
nonadditivity increases the clustering over the RPM fluid and a
positive nonadditivity diminishes it. Summarizing, for the
  fixed values of {$|\Delta|$} used, we find, in agreement with
  Ref. \cite{Pastore1985}, that: 
(a) at high density and positive $\Delta$ we have more clustering than
in the additive model, (b) at high density and negative $\Delta$ we
have less clustering than in the additive model, (c) at low density
and positive $\Delta$ we have less clustering than in the additive
model, (d) at low densities and negative $\Delta$ we have more
clustering than in the additive model. {These points can be
  explained observing that a pair of unlike ions have a higher
  affinity with negative $\Delta$. Thus, in a bulk phase negative
  $\Delta$ favors etherocoordination. Clusters of a given number of ions
  tend to be smaller when $\Delta$ is negative. As a result, at low
  density (where excluded volume plays a small role), the extra
  affinity due to negative $\Delta$ enhances cluster formation. By
  contrast, at high densities, the increase in available volume
  from the resulting etherocoordination with negative $\Delta$ has an
  important role, reducing the density-driven imperative to form
  clusters in the negative $\Delta$ case. The same arguments in
  reverse explain the behavior of a system with positive
  nonadditivity where now homocoordination at high density is favored
  \cite{Pastore1985}.}   

To qualitatively reproduce the curves with non-percolating clusters we
can use the Tani and Henderson clustering analysis 
\cite{Tani1983,Fantoni2011,Fantoni2012} with an 
inter-cluster configurational partition function the one of an ideal
gas of clusters, in reduced units, $Z_\text{inter}\approx (V/\sigma^3)^{N_t}$,
where $N_t=\sum_{n=1}^{n_c} N_n$ is the total number of clusters and
we assume to have only clusters made of up to $n_c$ particles. Then
the equations for the equilibrium cluster concentrations are
\bq \label{Tani1}
\langle N_n\rangle/N&=&\lambda^nz^\text{intra}_n/\rho^*~,~~~n=1,2,\ldots,n_c~,\\
\label{Tani2}
1&=&\sum_{n=1}^{n_c} n\langle N_n\rangle/N~,
\eq
where $z_n^\text{intra}$ are the configurational intra-cluster partition
functions in reduced units {with $z_1^\text{intra}=2$} and
$\lambda(=\alpha\rho^*/2)$ is a 
Lagrange multiplier {to be determined by Eq. (\ref{Tani2})}. 
Moreover neglecting the excess internal energy of the clusters we can
approximate {$z^\text{intra}_{n}\approx
  (v_n/\sigma^3)^{(n-1)}2^n/n!$ where $v_n$ is the volume of an
  $n-$cluster. Assuming further the cluster to 
  be in a closed packed configuration we can
  approximate\footnote{Clearly a proper analysis of the 
cluster volume would itself require a MC simulation
\cite{Gillan1983}.} $v_n\approx
n\sigma^3/\sqrt{2}$. This simple approximation is temperature
independent and its usefulness is thereby quite limited.} 

We checked the size dependence of the curves shown in
Fig. \ref{fig:ncl-t0.1} and saw that when we have no percolating
clusters the curve was unaffected by a choice of an higher number of
particles (up to 5000), while the curve changed in presence
of percolating clusters. In this case we 
found that a common curve is given by $\langle N_x\rangle/N$ with
$x=n/N\in [0,1]$. Then, in order to satisfy the normalization
condition,     
$1=\sum_nn(\langle N_n\rangle/N)\approx\int dx\,xN^2(\langle
N_x\rangle/N)$, we must have $(\langle N_x\rangle/N^\prime)/(\langle
N_x\rangle/N^{\prime\prime})\approx (N^{\prime\prime}/N^\prime)^2$ for
two different sizes $N^\prime$ and $N^{\prime\prime}$.  

\begin{figure}[hbt]
\begin{center}
\includegraphics[width=7cm]{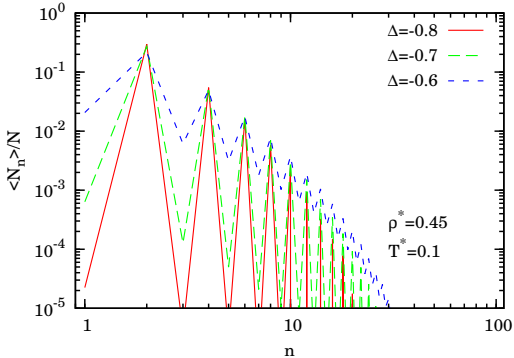}
\end{center}  
\caption{(color online). We show the clustering properties of the fluid at
$T^*=0.1$ and 
$\rho^*=0.45$ at various values of negative nonadditivity approaching
$-1$. $N_n$ are the number of  
clusters made of $n$ particles. We chose
$\delta=0.1$. In the MC simulations we used $N=100$
particles and a number of MCS$=5\times 10^7$.}
\label{fig:dip}
\end{figure}

In Fig. \ref{fig:dip} we show the clustering analysis for the fluid
with $\Delta$ approaching $-1$ at $T^*=0.1$ and $\rho^*=0.45$. We see
how letting $\Delta$ approach $-1$ this stabilizes {small}
neutrally charged clusters and lowers the degree of dissociation
{$\alpha=\langle N_1\rangle/N$}. The first stable 
cluster is the dipole: the ``overlap'' of a positive and a negative
sphere. This are dipoles of moment $qr_{12}$ with $r_{12}<\sigma
(1+\Delta+\delta)$ which may lack a gas-liquid criticality
\cite{Rovigatti2011}. We clearly have a transition from a conducting 
to an insulating phase as $\Delta$ goes from $0$ to $-1$. We expect
that in the limiting case of $\Delta=-1$ the {system} we
obtain is the  
{neutral HS fluid of half the density}. This is confirmed
by a comparison of the like radial distribution functions with the one
of the neutral HS
even if the $\Delta=-1$ fluid simulation rapidly slows down
into the frozen configuration of the overlapping anions and cations. In
order to overcome this problem one should alternate single particle
moves with neutrally charged $2-$cluster moves. 

In order to
qualitatively reproduce the curve of Fig. \ref{fig:dip} we need to use
Eqs. (\ref{Tani1})-(\ref{Tani2}) with $z_{n}^\text{intra}=\sum_{s=0}^n
z_{s,n-s}^\text{intra}$ {where $z_{s,t}^\text{intra}$ are the
  configurational intra-cluster partition functions of a cluster made
  of $s$ anions and $t$ cations,} 
\bq \nonumber
z_{s,t}^\text{intra}&=&\frac{1}{s!t!}\int_{\Omega_{s,t}}
\frac{d\rr_2\ldots
d\rr_{s+t}}{\sigma^{3(s+t-1)}}\times\\ \label{zintra0}
&&e^{-\beta\sum_{\mu>\nu=1}^{s+t}\phi_{i_\mu
    j_\nu}(r_{\mu\nu})}\\ \label{zintra}
&\approx&\frac{(s+t)^{(s+t-1)}}{s!t!}(K/K_0)^{\min\{s,t\}}~,\\
K/K_0&=&\int_{\sigma(1+\Delta)}^{\lambda_B/2}r^2e^{+\lambda_B/r}\,dr/
\int_{\sigma(1+\Delta)}^{\lambda_B/2}r^2\,dr~,
\eq
where the configurational integral goes only over the relative
positions and it covers the region $\Omega_{s,t}$ of $s$ anions
clusters configuration space, $\lambda_B=\sigma/T^*$ is the
  Bjerrum length, Roman indeces denote the particle species, Greek
  indeces denote the particle labels, a Roman index 
  with a Greek subindex denotes the species of the particle
  corresponding to the Greek subindex, and $\rr_{\mu\nu}$ denotes the
  separation vector between particle $\mu$ and particle $\nu$. 
  Eq. (\ref{zintra}) is justified as follows. Let us call the
  anions $i_-=1_-,\ldots,s_-$ and the cations
  $j_+=1_+,\ldots,t_+$. From Eq. (\ref{zintra0}) follows 
\bq \nonumber
z_{t,t}^\text{intra}&=&\frac{1}{t!^2}\frac{1}{\sigma^{3(2t-1)}}\int_{\Omega_{t,t}}
\prod_{l=2}^td\rr_{1_+l_-}\prod_{k=1}^td\rr_{k_+k_-}\\ \nonumber
&&\times\prod_{i>j=1}^te^{-2\lambda_B/r_{i_+j_+}}\prod_{i,j=1}^te^{+\lambda_B/r_{i_+j_-}}\\ 
\nonumber 
&\approx&\frac{1}{t!^2}\frac{1}{\sigma^{3(2t-1)}}\int_{\Omega_{t,t}}
\prod_{l=2}^td\rr_{1_+l_-}\prod_{k=1}^td\rr_{k_+k_-}\\ 
&&\times\prod_{i,j=1}^te^{+\lambda_B/r_{i_+j_-}}~,
\eq
where we approximated $e^{-\lambda_B/r}\approx 1$ which is justified
at high $T^*<1/2(1+\Delta)$ or low $\lambda_B$. Now we observe that
for example $r_{1_+2_-}=|\rr_{1_+1_-}+\rr_{1_-2_-}|$ with
$r_{1_-2_-}>\sigma$ and $e^{+\lambda_B/r_{1_+2_-}}\approx 1$. So that for
negative nonadditivity we can further approximate   
\bq \nonumber
z_{t,t}^\text{intra}&\approx&\frac{1}{t!^2}\frac{1}{\sigma^{3(2t-1)}}\int_{\Omega_{t,t}}
\prod_{l=2}^td\rr_{1_+l_-}\prod_{k=1}^td\rr_{k_+k_-}\\ \nonumber
&&\times\prod_{i,j=1}^te^{+\lambda_B/r_{i_+j_-}}\\ \nonumber
&\approx&\frac{1}{t!^2}\frac{1}{\sigma^{3(2t-1)}}\int_{\Omega_{t,t}}
\prod_{l=2}^td\rr_{1_+l_-}\prod_{k=1}^td\rr_{k_+k_-}\\ \nonumber
&&\times\prod_{i=1}^te^{+\lambda_B/r_{i_+i_-}}\\
&\stackrel{\mbox{\normalsize  $\propto$}}{\sim}&
\frac{(2t)^{(2t-1)}}{t!^2}(K/K_0)^t~,
\eq
where the factor $(2t)^{(2t-1)}$ takes into account the volume of
$\Omega_{t,t}$. Using the same chain of approximations we reach
Eq. (\ref{zintra}). {We immediately see how
  $z_{1,1}^\text{intra}\propto K/\sigma^3$ becomes bigger and bigger as
  $\Delta\to -1$ and the same holds for all the $z_{k,k}^\text{intra}$
  which clearly dominate over all the others $z_{s,t}^\text{intra}$ with
  $s\neq t$. And this qualitatively explains Fig. \ref{fig:dip}.} 

%

\label{sec:binodal}

Sufficiently close to the critical point we determined the qualitative
change in the behavior of the gas-liquid coexistence region  
by switching on a negative or a positive nonadditivity. To this
aim we divided the simulation box into $m^3$ cubes of side $\ell=L/m$ and
registered the density inside each cell
{$\rho_i={\cal N}_i/\ell^3$}, where {${\cal N}_i$} is
the number of particles inside the 
$i$th cell so that {$\sum_{i=1}^{m^3}{\cal N}_i=N$}. Then we
calculated the density 
distribution function $P_m(\rho)=\sum_{i=1}^{m^3}P_m(\rho_i)/m^3$
\cite{Rovere1988,Rovere1990}, where $P_m(\rho_i)$ is the distribution
function for the $i$th cell. With $\int P_m(\rho)\, d\rho=1$. Above
the critical temperature the density probability distribution function
can be described by a Gaussian centered at the simulation density
whereas below it becomes bimodal with two peaks one centered at the
gas density and one at the liquid density.

We start from an initial configuration of particles of random species
placed on a simple cubic lattice. We equilibrate (melt) the fluid for
$10^6$ MCS$/$particle. We then sampled the distribution function
every $10$ MCS. To {enhance the efficiency of the
  determination of the cell density distribution}, every $10$ MCS, we
choose the subdivision of the simulation box 
in cells with a random displacement $\rr=(r_x,r_y,r_z)$ with $r_x,r_y,r_z\in
[0,L]$. And we measured the distribution function on runs of $10^6$
MCS$/$particle.   

\begin{figure}[hbt]
\begin{center}
\includegraphics[width=7cm]{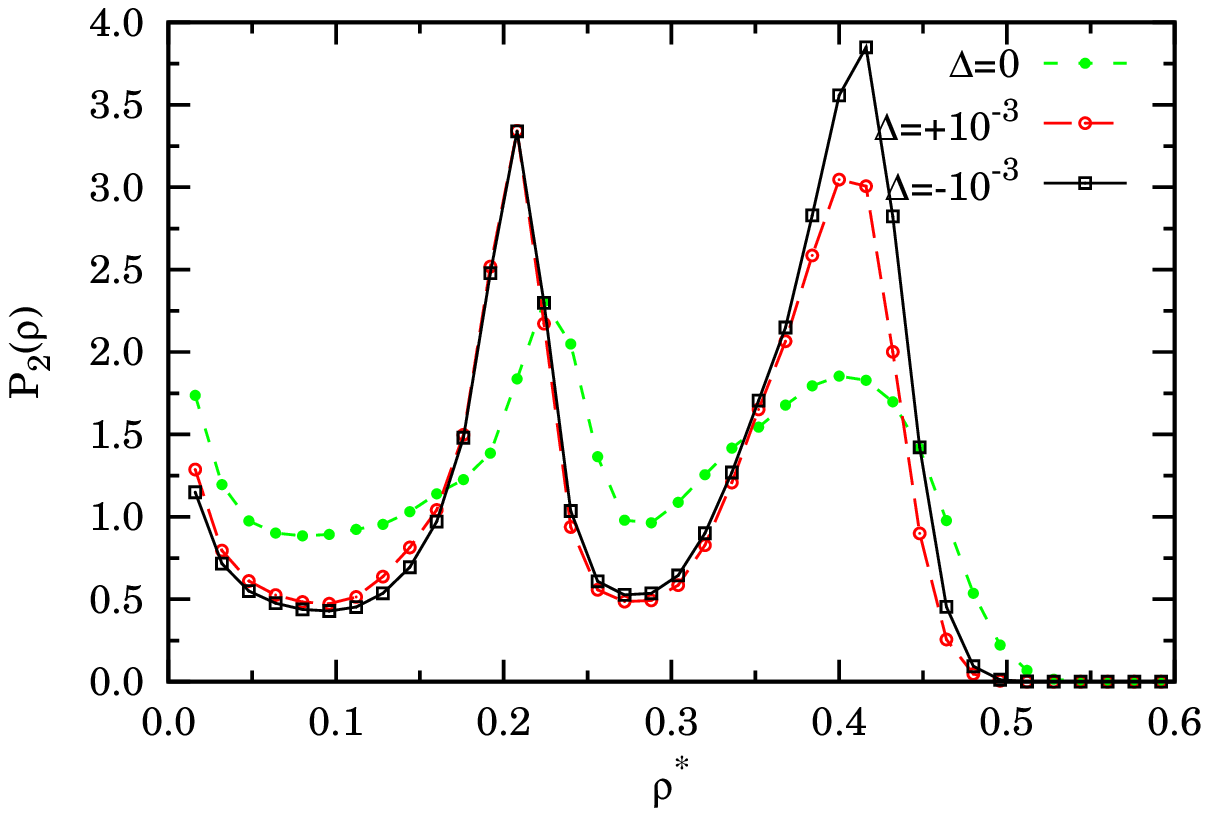}
\includegraphics[width=7cm]{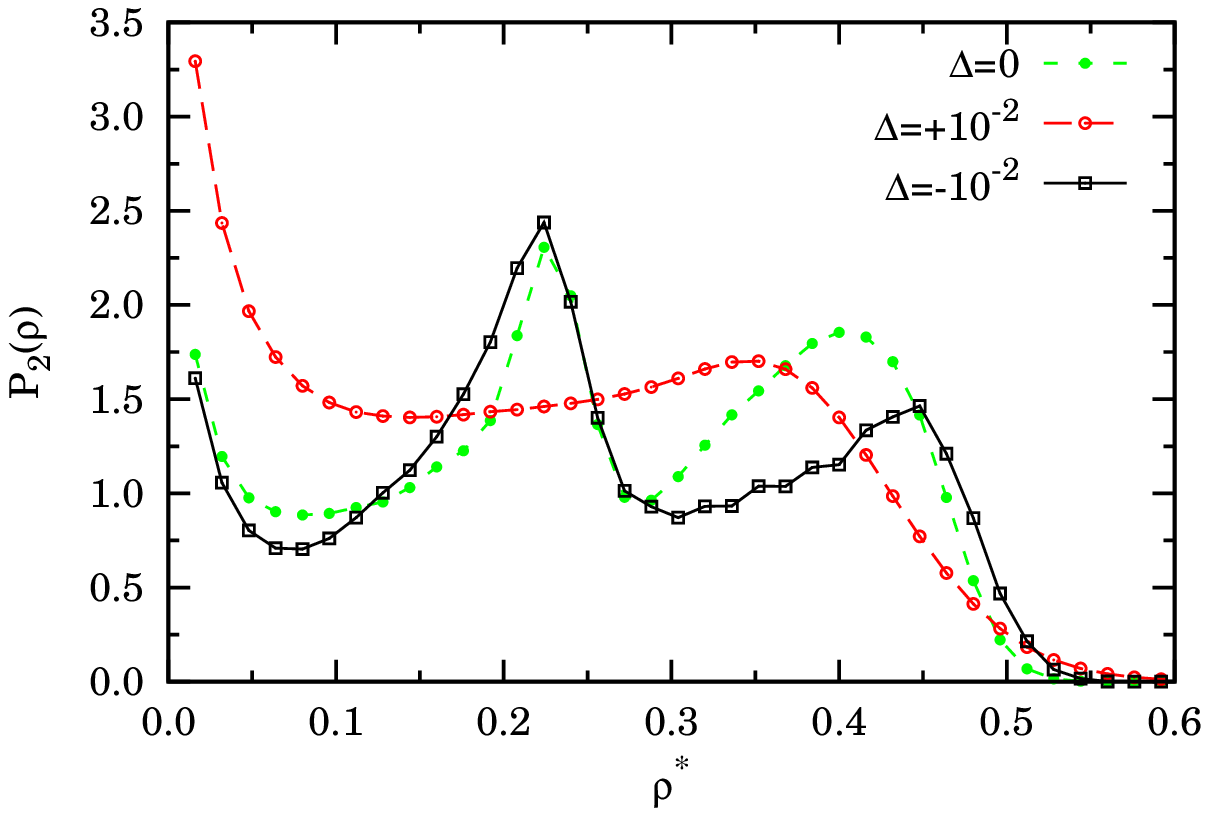}
\includegraphics[width=7cm]{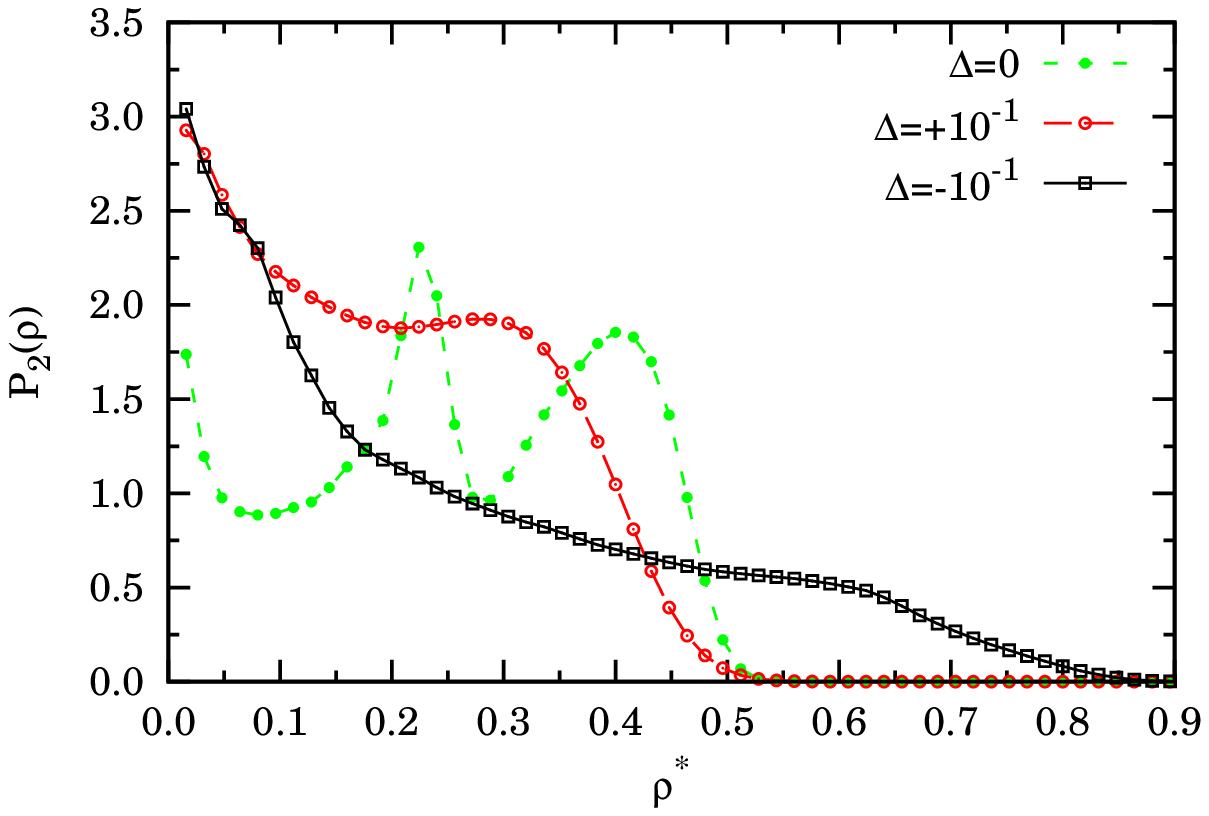}
\end{center}  
\caption{(color online). Cell density distribution function for the fluid at
  $T^*=0.02, \rho^*=0.2$ and $\Delta=0,\pm {\cal D}$ with
  {${\cal D}=10^{-3},10^{-2},10^{-1}$}. We used $N=100$
  and $m=2$ with $10^6$ MCS$/$particle.}   
\label{fig:bim-r0.2-t0.02}
\end{figure}

Choosing $m=2$ and $N=100$ the results for the fluid at a temperature
$T^*=0.02, \rho^*=0.2$, well within the coexistence region of the pure
RPM fluid, and $\Delta=0,\pm {\cal D}$ with {${\cal D}=10^{-1},
10^{-2},10^{-3}$} are shown in Fig. \ref{fig:bim-r0.2-t0.02}. In this 
case the minimum density that can be registered is $1/\ell^3=0.2\times
8/100=0.016$. We see that the pure RPM fluid shows a density
distribution function which has three peaks with the first peak, which
lies below the minimum density, at approximately the low density of
the gas phase, the second peak at the simulations density $\rho^*=0.2$
which is due to the fact that the fluid
develops surfaces between the gas and the liquid phase
\cite{Smit1989}, and the third peak at approximately the high density 
of the liquid phase. {We see from the figure that increasing
  ${\cal D}$ the middle peak is lost first in the positive additive
  model and then in {the negative} nonadditive
  models. Moreover for the biggest ${\cal D}$ the peak of the liquid
  phase is barely visible. This may be due to the fact that one had to
  choose a {proper} simulation density {closer to
    the density of the liquid \cite{Rovere1988,Rovere1990}}.} We 
clearly see how this analysis works like a ``microscope'' on the
degree of nonadditivity predicting an increase(decrease) of the
coexistence region for small negative(positive)
nonadditivity. {This behavior can be explained as follows. 
Positive nonadditivity increases the effective excluded volume of
ions, thereby reducing the density of the liquid phase, and negative
nonadditivity does the opposite.}

{We believe that our results could be relevant for the
interpretation of experimental work on the phase diagrams of
room temperature ionic liquids \cite{Saracsan2006}. In these
experimental systems the liquid-liquid binodals shifted above and
below the one of the pure RPM are observed depending on the kind of
solvent. If on the one hand this can be ascribed to the different
dielectric constant of the solvent \cite{Kleemeier1999} on the other
hand it is clear that depending on the kind of solvent used the
anion-cation contact-pairing affinity may vary \cite{Kalcher2010} 
and
thus the different experimental ionic liquids considered should be
more correctly described by comparison not just with the pure RPM but
with the more realistic primitive model with the addition of either a
positive or negative size nonadditivity.}

\label{sec:conclusions}

In conclusion, we have performed for the first time a MC
simulations study of the vapor phase of the RPM with nonadditive
diameters, with particular emphasis on its clustering properties. A 
density distribution function analysis shows how the gas-liquid
coexistence region evolves by switching on the nonadditivity. 
A negative nonadditivity tends to enlarge the coexistence
region while a positive one to shrink it.

From the cluster analysis we where able to distinguish between two
kind of behaviors for the cluster concentrations. When we are below
the percolation threshold the curves for the
cluster concentrations as a function of the cluster size are
independent of the number of particles used in the simulation and
can be {qualitatively explained by a simple clustering
theory where one approximates the clusters 
to form an ideal gas and the $n-$cluster as formed by $n$
non-interacting particles}, for not too small  
density or nonadditivity. When we are above the
percolation threshold the curves depend on the number of particles
used in the simulation and obey a simple scaling relationship. 

At low density, the negative nonadditive fluid has higher clustering
than the pure RPM whereas at high densities {the positive 
nonadditive fluid has a greater degree of clustering}. The positive
nonadditive fluid is the first one to reach the percolating
clusters upon an increase of the density. This is due to the less
space available to the ions, for a given density, for positive
nonadditivity {and to the frustrated tendency to segregation
  of like particles at high density}. A negative nonadditivity tends
to greatly enhance the 
formation of the neutrally charged clusters, starting with the 
dipole, {as can be predicted} from the {simple}
clustering theory refined at the 
intra-cluster level. Traces of these features can also be
read from an analysis of the partial radial distribution function
and structure factors, which will be presented elsewhere. 

In parallel with the density distribution function analysis we are
currently planning to perform a Gibbs ensemble MC study of the 
gas-liquid binodal to establish more accurately the dependence on the
nonadditivity parameter. 
    
We hope that the present study could foster additional theoretical and
computational studies as well as  experimental realizations of these 
simple but rich fluids.
\acknowledgments
R.F. would like to acknowledge the use of the computational facilities
of CINECA through the ISCRA call. {Both authors would like to
thank the referee for useful comments.}


\end{document}